\documentclass[useAMS,usenatbib]{mn2e}

\usepackage{amsfonts,amsmath,amssymb,mathrsfs}
\usepackage{graphicx,epsf,epsfig}
\usepackage{bm}
\usepackage{longtable}
\usepackage[usenames,dvipsnames]{xcolor}
\usepackage{multirow}

\usepackage{graphicx}
\usepackage{hyperref}
\usepackage{times}
\usepackage{color}
\usepackage{breakurl}
\usepackage{color}
\usepackage{mathrsfs}

\def\be{\begin{equation}}
\def\ee{\end{equation}}
\def\bea{\begin{eqnarray}}
\def\eea{\end{eqnarray}}

\title[Nonvalidity of $I$-Love-$Q$ Relations for Hot White Dwarf Stars]{Nonvalidity of $I$-Love-$Q$ Relations for Hot White Dwarf Stars} 

\author[K.~Boshkayev and H.~Quevedo]{K.~Boshkayev$^{1,2}$ and H.~Quevedo$^{1,2,3}$\\
$^1$NNLOT, Al-Farabi Kazakh National University, Al-Farabi av. 71, 050040 Almaty, Kazakhstan\\ 
$^2$Dipartimento di Fisica and ICRA, Universit\`a di Roma "La Sapienza",  Piazzale Aldo Moro 5, I-00185 Roma, Italy\\
$^3$Instituto de Ciencias Nucleares, Universidad Nacional Aut\'onoma de M\'exico, AP 70543, M\'exico, DF 04510, Mexico}

\begin{document}
\date{\today}
\maketitle

\begin{abstract}
The equilibrium configurations of uniformly rotating white dwarfs at finite temperatures are investigated, exploiting the Chandrasekhar equation of state for different isothermal cores. The Hartle-Thorne formalism is applied to construct white dwarf configurations in the framework of Newtonian physics. The equations of structure are considered in the slow rotation approximation and all basic parameters of rotating hot white dwarfs are computed to test the so-called moment of inertia, tidal Love number and quadrupole moment ($I$-Love-$Q$) relations. It is shown that even within the same equation of state the $I$-Love-$Q$ relations are not universal for white dwarfs at finite temperatures.
\end{abstract}


\begin{keywords}
Hartle-Thorne formalism, equilibrium configurations, stars: white dwarfs, $I$-Love-$Q$ relations, finite temperatures
\end{keywords}


\section{Introduction}
\label{intro}
The no-hair theorems for black holes state that in general relativity only a finite number of multipole moments are necessary to describe black holes, namely, mass, charge and angular momentum \citep{heu96}. It is believed that during the gravitational collapse of an arbitrary mass distribution, higher multipoles disappear as a result of the emission of gravitational waves. One would then expect that other compact objects like neutron stars (NSs) are characterized in general by an infinite number of multipoles. However, some recent intriguing results seem to indicate that compact objects other than black holes are also characterized by a finite number of multipoles. 

In fact, the $I-$Love$-Q$ relation states that there exists a connection between the moment of  inertia, quadrupole moment and the Love numbers, which in compact objects measure their rigidity and  shape response to tidal forces. This relation is valid independently of the equations of state (EoSs) used to describe relativistic compact objects such as NSs and quark stars, if the slow rotation approximation is assumed in the framework of the Hartle-Thorne formalism \citep{yagi2013,maselli2013}. Similar approximate relations among multipole moments for NSs {have been also} investigated in the case of both slow and rapid rotation regimes \citep{yagi2014b,stein2014}.

The $I$-Love-$Q$ and $I$-Love relations, respectively, were investigated for incompressible and realistic stars; it was shown that the EoS-independent behaviour of the $I$-Love-$Q$ relation can be attributed to its incompressible limit \citep{sham2015, chan2015}.
Moreover, these relations were also calculated by \citet{pani2015} for exotic objects such as thin-shell gravastars at zero  temperature,  without assessing their validity. The nonvalidity for the thin-shell gravastars was shown for different EoSs 
\citep{uchikata2016}. Nonetheless, it was established that in gravastars these relations  posses distinct features from the ones of NSs and QSs. 

More recently, the validity of the $I-$Love-$Q$ relation was proven also in the case of dark stars by \citet{maselli2017} and white dwarfs (WDs) by \citet{bosh2017} at zero temperature. In the case of WDs, the Hartle-Thorne formalism was implemented in Newtonian physics to integrate the field equations together with the condition of hydrostatic equilibrium.

In the present work, we consider an additional important aspect of the internal structure of WDs, namely, their thermodynamic behavior. In particular, we analyze in detail the effects that follow from considering finite temperatures in the EoS. We will use the Hartle-Thorne formalism the validity of which has been well established in the derivation of all physically relevant quantities of rigidly rotating relativistic and classical objects \citep{Hartle,Hartle2,bosh2016}. The parameters describing the structure play a paramount role in the investigation of the stability and the lifespan of WDs \citep[see e.g.]{boshijmpe, boshizzo, brrs2013, boshJKPS2014, boshmg13, rueda2013}.

It has been shown that for massive white dwarfs close to the Chandrasekhar mass limit the effects of finite temperatures are negligible. However, for the observed low-mass white dwarfs the effects are crucial \citep{carvalho2014}. From the astrophysical point of view it is hard to measure the radius of a star with respect to its mass and temperature. Hence, if we know the mass and temperature of a WD, we can theoretically calculate its radius as it will be always different from the cold (degenerate) case 
\citep{carvalho2014, carvalho2018}. Therefore, we study here the effects that rotation along with temperature cause on the structure of WDs; first, we consider the main physical parameters of WDs and study their dependence on the density of the star for different temperature values. This allows us to investigate in detail the $I$-$Q$, $I$-Love and Love-$Q$ relations,  and to demonstrate that they are not universal. The temperature effects are sufficient to break down the universality of the $I-$Love$-Q$ relations. This is shown by integrating numerically the structure equations for slowly rotating WDs with the Chandrasekhar EoS 
\citep{Chandrasekhar1931,Chandrasekhar1939,rrrx} at different temperatures.


\section{Formalism and stability criteria for rotating hot white dwarfs}

A general relativistic analysis of the hydrodynamic equilibrium of WDs has established that relativistic effects lead only to small perturbations of Newtonian gravity \citep{mn14}; consequently, the essential physical features of WDs can be studied by using Newton's theory. If in addition we use the Hartle-Thorne formalism to analyze  perturbatively the structural equations, as proposed recently by \citet{bosh2017}, it is possible to explore in detail the behavior of all the relevant physical quantities. The main idea consists in solving Newton's equation 
\be
\nabla^2 \Phi = 4 \pi G \rho\ , 
\label{new}
\ee
and the equilibrium condition
\be
\frac{d p }{dr } = - \rho \frac{G M}{r^2} \ ,\qquad \frac{dM}{dr} = 4 \pi r^2 \rho\ ,
\label{equi}
\ee
perturbatively by expanding the radial coordinate as $r=R + \xi$, where $R$ is the radial coordinate for a spherical configuration and the function $\xi(R,\theta)$ takes into account the deviations from spherical symmetry due to the rotation of the star. All the relevant quantities such as the total mass $M$, equatorial radius $R_e$,  moment of inertia $I$, angular momentum $J$, quadrupole moment $Q$, etc. are then Taylor expanded up to the second order in the angular velocity. Within this approximation, due to an appropriate choice of $\xi$, the density $\rho$ and pressure $p$ can be considered as non affected by the rotation of the star.  The structural equations (\ref{new}) and (\ref{equi}) can then be integrated numerically to obtain all the relevant quantities in the desired approximation. 

For the analysis of the structural equations it is convenient to introduce the 
Keplerian angular velocity
\begin{equation}\label{eq:omegaK}
\Omega_{Kep}=\sqrt{\frac{GM}{R^3_e}} \ ,
\end{equation}
because it allows us to calculate all the key parameters at the mass-shedding limit, and to determine the stability region inside which rotating configurations can exist \citep{brrs2013}.

Finally, the inverse $\beta$-decay instability determines the critical density which, in turn, defines the onset of instability for a WD to collapse into a neutron star. For the Chandrasekhar EoS we adopt $\rho_{crit}=1.37\times10^{11}$ g/cm$^3$. The inverse $\beta$-decay instability is crucial both for static and rotating configurations. It represents one of the boundaries of the stability region for  rotating WDs \citep{brrs2013, bosh2016, bosh2017}.
According to \citet{carvalho2014}, the occurrence of the inverse $\beta$-decay instability is not affected by the presence of temperature, i. e., it is the same as in the degenerate approximation. This  is related to the fact that the effects of temperature are negligible in the higher density regime. For the sake of generality, all computations are performed for central densities up to 
$10^{12}$ g/cm$^3$.


\section {Equations of state for white dwarfs}

We will use the simplest EoS for WD matter that correctly describes its main physical properties, namely, the Chandrasekhar EoS  \citep{carvalho2014, bosh2016b}. Then, the total pressure is due to the pressure of electrons ${P}_{e}$, because the pressure of positive ions ${P}_{N}$ (naked nuclei) is insignificant, whereas the energy density is due to the energy density of nuclei ${\cal E}_{N}$ as the energy density of the degenerate electrons ${\cal E}_{e}$ is negligibly small. Thus, the Chandrasekhar EoS is given by
\be
\label{eq:ECh0}
{\cal E}_{Ch}  =  {\cal E}_{N}+{\cal E}_{e}\approx {\cal E}_{N}, 
\ee
\be
\label{eq:PCh0}
{P}_{Ch}  =  {P}_{N}+{P}_{e}\approx{P}_{e}.
\ee
Hence the energy density of the nuclei is given by
\be
{\cal E}_{N}=\frac{A}{Z} M_{u}c^{2}n_{e}
\ee
where $A$ is the average atomic weight, $Z$ is the number of protons, $M_{u}=1.6604\times10^{-24}$ g is the unified atomic mass, $c$ is the speed of light and $n_{e}$ is the electron number density. In the following analysis, we will assume the average molecular weight $\mu=A/Z=2$.
In general, the electron number density follows from the Fermi-Dirac statistics, and is determined by 
\be\label{eq:ne}
n_e=\frac{2}{(2\pi \hbar)^{3}} \int_0^\infty \frac{4\pi p^{2} dp}
{\exp\left[\frac{\tilde{E}(p)- \tilde{\mu}_{e}(p)}{k_B T}\right] +1} ,
\ee
where $k_B$ is the Boltzmann constant, $\tilde{\mu}_{e}$ is the electron chemical potential without the rest-mass, 
and $\tilde{E}(p)= \sqrt{c^2 p^2+m^2_e c^4} - m_{e}c^2$, with $p$ and $m_e$ being the electron momentum and rest-mass, respectively, \citep{carvalho2014}.

The electron number (\ref{eq:ne}) can be written in an alternative form as
\be
n_{e}=\frac{8\pi \sqrt{2}}{(2 \pi \hbar)^3} m^3 c^3 \beta^{3/2} \left[ F_{1/2} (\eta,\beta) + \beta F_{3/2} (\eta,\beta) \right],
\label{eq:ne2}
\ee
where 
\begin{equation}
F_{k} (\eta,\beta)=\int_{0} ^{\infty} \frac{t^k \sqrt{1+(\beta/2)t}}{1+ e^{t-\eta}}\, dt
\label{eq:Fk1}
\end{equation}
is the relativistic Fermi-Dirac integral, $\eta=\tilde{\mu}_e/(k_B T)$, $t=\tilde{E}(p)/(k_{B}T)$ and $\beta=k_{B}T/(m_e c^2)$ are the degeneracy parameters. 
Consequently, the total electron pressure for $T\neq0$ K is given by 
\begin{align}\label{eq:Pe2}
P_e=\frac{2^{3/2}}{3 \pi^2 \hbar^3} m_{e}^4 c^5 \beta^{5/2} \left[ F_{3/2} (\eta,\beta)\right.+ \left. \frac{\beta}{2} F_{5/2} (\eta,\beta) \right].
\end{align}

When $T=0$, for a degenerate electron gas we find from Eq.~(\ref{eq:ne}) that
\begin{equation}
n_e=\int_{0}^{P^F_e}\frac{2}{(2\pi\hbar)^3}d^{3}p=\frac{(P^F_e)^3}{3{\pi}^2{\hbar}^3}= \frac{(m_e c)^3}{3{\pi}^2{\hbar}^3}x_e^3 .
\end{equation}
Thus, the total electron pressure is
\be
\begin{split}
P_e = \frac{1}{3} \frac{2}{(2 \pi \hbar)^3} \int_0^{P^F_e} \frac{c^2 p^2}{\sqrt{c^2 p^2+m^2_e c^4}} 4 \pi p^2 dp \qquad \\ = \frac{m^4_e c^5}{8 \pi^2 \hbar^3}\left[x_e \sqrt{1+x^2_e}\left(\frac{2x^2_e}{3}-1\right)+{\rm arcsinh}(x_e)\right] \label{eq:eos2}\, ,
\end{split}
\ee
where $x_e = P^F_e/(m_e c)$ is the dimensionless Fermi momentum.


\section{Results and discussion}

\begin{figure}
\centerline{\includegraphics[width=8cm]{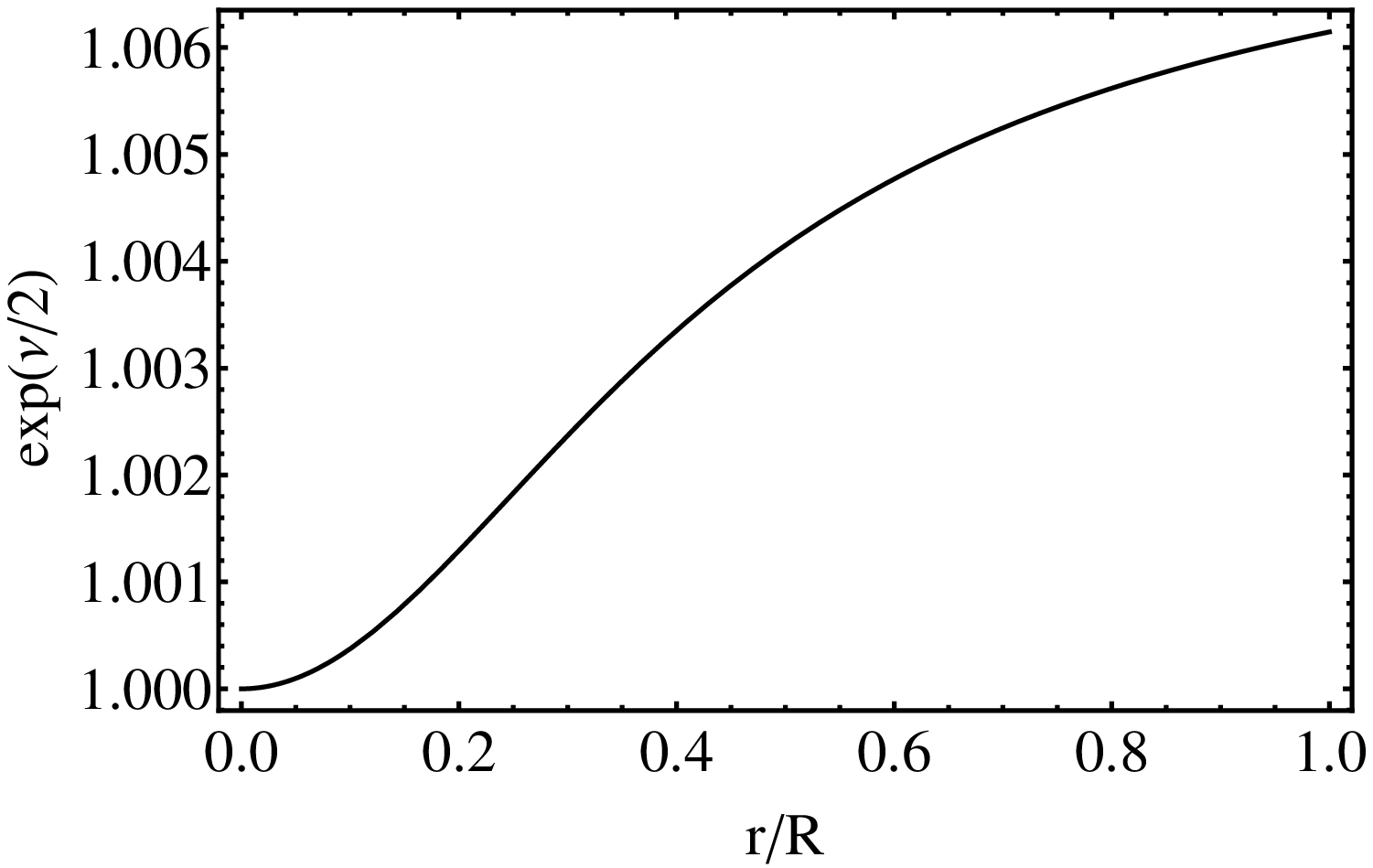}}
\caption{$\exp(\nu/2)$ as a function of the radial distance for a zero temperature white dwarf of mass $M$= 1.44$M_{\odot}$ and radius $R$=1000 km.}
\label{fig:expr1}
\end{figure}

For the sake of simplicity, throughout the paper we use a uniform temperature profile for isothermal cores of WDs, i.e. WDs without outer envelop (atmosphere).  The atmosphere serves as an isolator and its effect on the structure of WDs can be neglected in this approximation. In order to justify a constant temperature profile within the core, we considered the  \cite{tolman1930}  equilibrium condition for hot relativistic static stars, which is given by $T/u^t = constant$, where $T$ is the local temperature and $u^t$  is the zero-component of the four-velocity. In the case of a static star, $u^t = 1/\sqrt{g_{tt}}$,  from which one obtains the known Tolman law: 
$\sqrt{g_{tt}}T = constant$. So, for the usual spherically symmetric metric: $\exp(\nu/2)T = constant$.
In the classical limit $\exp(\nu/2) \approx 1-\Phi/c^2$, where $\Phi= \Phi(r)$ is the internal Newtonian gravitational potential found from Eq.~(\ref{new})  and $c$ is the speed of light in vacuum. We constructed $\exp(\nu/2)$ as a function of $r/R$. We then selected a white dwarf with mass 1.44$M_{\odot}$ and radius 1000 km, as an example. As one can see from Fig.~\ref{fig:expr1}, the function $\exp(\nu/2)$ changes slightly from the center to the surface of a white dwarf core. So, $\exp(\nu/2)$ changes less than 1$\%$ from the center to the surface of the isothermal core. This is the main argument to adopt the constant temperature profile.

\begin{figure}
\centerline{\includegraphics[width=8cm]{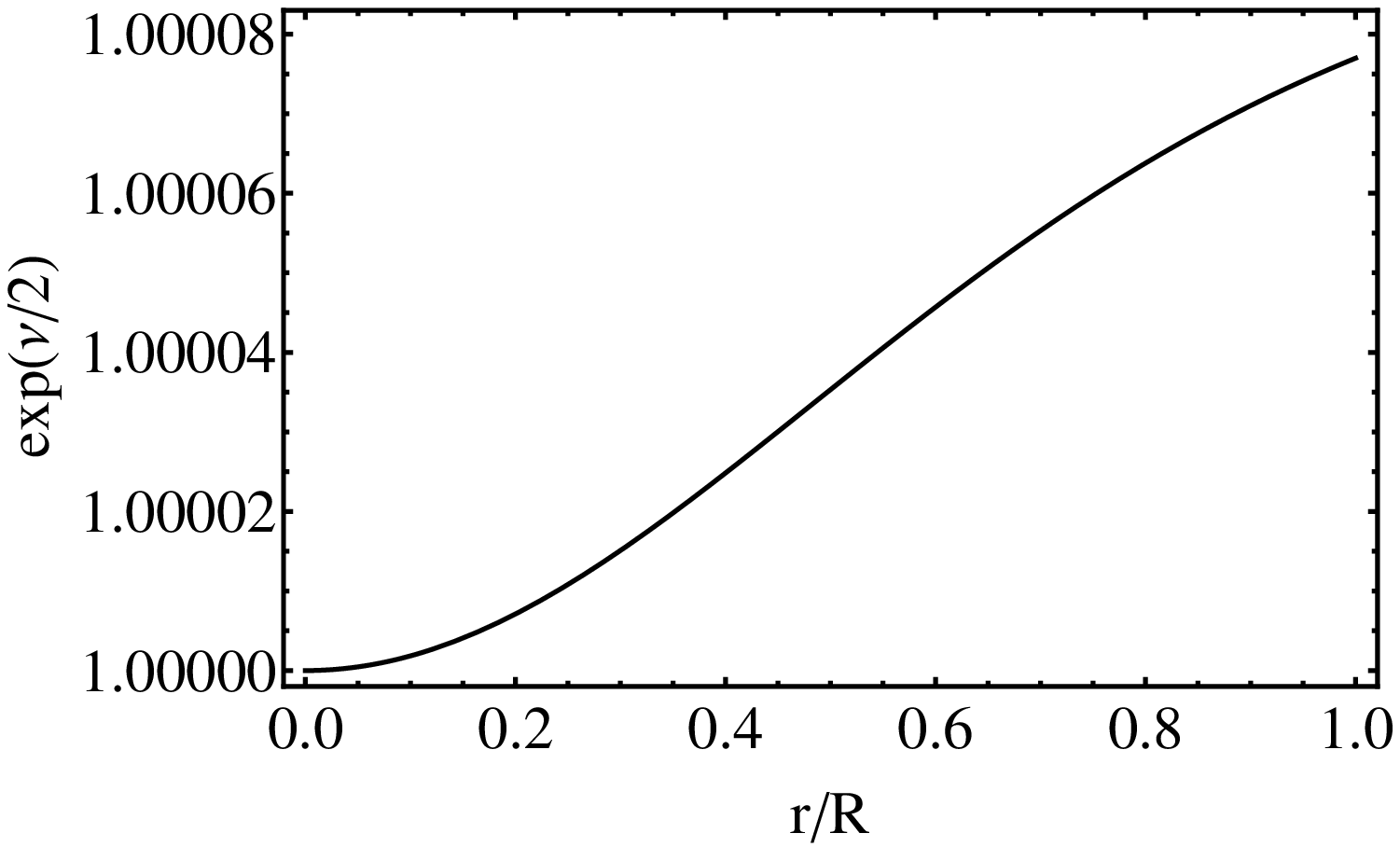}}
\caption{$\exp(\nu/2)$ as a function of the radial distance for a zero temperature white dwarf of mass $M$=0.4$M_{\odot}$ and radius $R$=10952 km.}
\label{fig:expr2}
\end{figure}

One can  calculate $\exp(\nu/2)$ also for a low mass white dwarf with  mass $0.4M_{\odot}$ and radius 10952 km. 
In Fig.~\ref{fig:expr2}, we see that $\exp(\nu/2)$ changes even less than in the previous case. Hence, for the cores of WDs the constant temperature profile is a safe assumption. A further generalization of the Tolman condition for slowly rotating stars is given by 
\cite{belvedere2014}. Even in that general case the change of function $\exp(\nu/2)$ turned out to be negligible for WDs.

\begin{figure}
\centerline{\includegraphics[width=8.5cm]{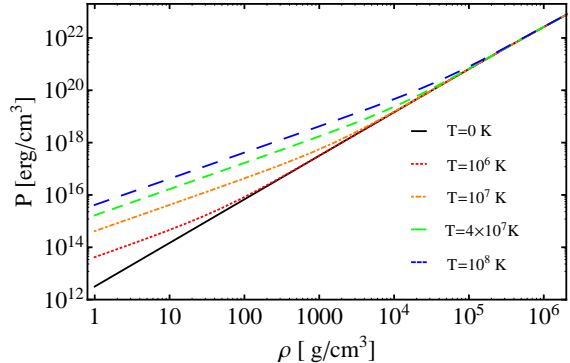}}
\caption{Total pressure as a function of the mass density for selected temperatures in the range $T=\left[0,10^8\right]$K (colour online).}
\label{fig:Prho}
\end{figure}

In Fig.~\ref{fig:Prho}, we plot the total pressure Eq.~(\ref{eq:PCh0}) as a function of the total density Eq.~(\ref{eq:ECh0}) for some selected temperatures. We conclude that  the effects of temperature are essential only in the range of small densities.

\begin{figure}
\includegraphics[width=8.5cm]{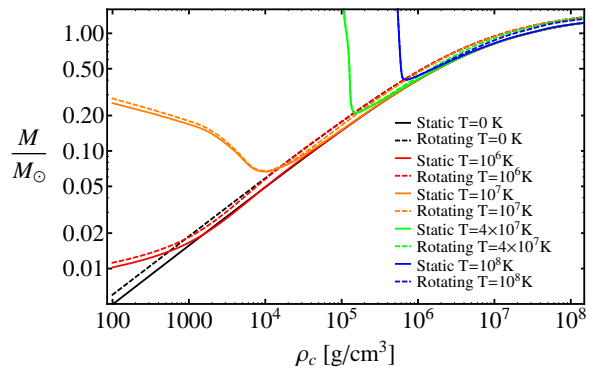} 
\caption{Mass versus central density (colour online).}\label{fig:Mrho} 
\end{figure}

In general, we assume that the slow-rotation approximation can be applied to any realistic star with a Keplerian angular velocity. Indeed, \citet*{Hartle2} in their pioneering article used this approximation to investigate the effects and deviations produced by rotation starting from massive non-compact stars to neutron stars. The general conditions in the slow-rotation regime are that 
the velocities of particles at the equatorial plane of the star must be non-relativistic  and, of course, that the fractional changes of density, pressure, mass, radius, gravitational potential etc., due to rotation must be smaller than in the static case. 
However, the most practical condition to check the validity of the slow rotation approximation for WDs would be to compare 
the mass-radius relations at the mass shedding limit within the slow-rotation approximation with the resuls obtained by using 
exact numerical computations. Unfortunately, to our knowledge, for white dwarfs this problem has not been considered yet. Some analysis of the validity of the slow approximation for WDs were performed by \citet[see][Appendix D, Fig. 9] {brrs2013}. Here we employ the Keplerian velocity to set upper bounds for all physical quantities as their realistic values will be between static and maximally rotating configurations. Hence, by solving the structure equations, we construct all necessary relations along the mass shedding sequence with angular velocity $\Omega_{Kep}$.

In Fig.~\ref{fig:Mrho}, the static and rotating mass of a WD is shown as a function of the central density and temperature. 
Our results show that in general  rotating WDs have larger masses than their static counterparts. Due to the choice of the scale the green and blue curves look sudden and sharp, but in reality they are not so abrupt. The curves look sharper with respect to colder white dwarfs, because of the pronounced effects of higher temperatures.

\begin{figure}
\includegraphics[width=8.5cm]{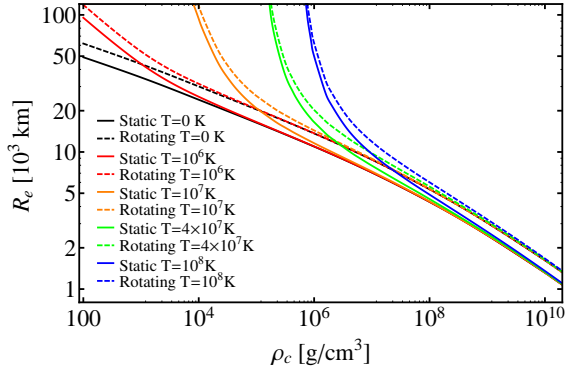} 
\caption{ Radius versus central density (colour online).}\label{fig:Rrho} 
\end{figure}
Fig.~\ref{fig:Rrho} shows the equatorial radius as a function of the central density and temperature for both rotating and static WDs.  
The plots show that hot WDs possess larger radius than cold ones. For increasing central densities, WDs become more gravitationally bound and spherical.

\begin{figure}
\includegraphics[width=8.5cm]{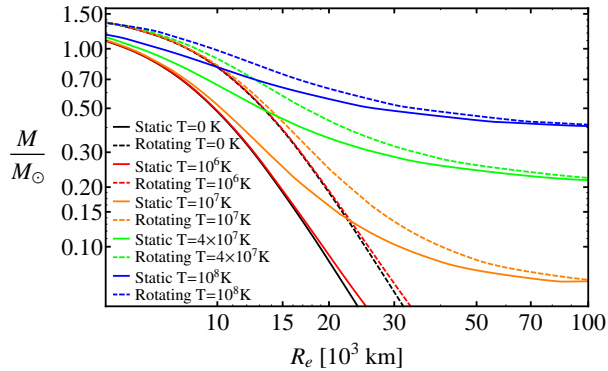} 
\caption{ Mass versus radius (colour online).}\label{fig:MR} 
\end{figure}
Fig.~\ref{fig:MR}  shows the mass and equatorial radius relation.  Here one can see that the mass-radius relation significantly diverges from the degenerate case especially for lower masses and larger radii, depending on the value of the core temperature. The relationship between the core temperature and observed effective surface temperature is given via the so-called Koester relation \citep{carvalho2014}. This explains the variety of observed WDs, according to the Sloan Digital Sky Survey Data Release. Indeed, nowadays we have data for more than thirty two thousand WDs and all of them have diverse characteristics \citep{kepler1, keplerscience, kepler2, koester2, Tremblay2011, kepler2007}. It should be stressed that the scale of the mass is selected for the sake of generality. Indeed, so far observed WDs have masses larger than 0.1$M_\odot$.

\begin{figure}
\includegraphics[width=8 cm]{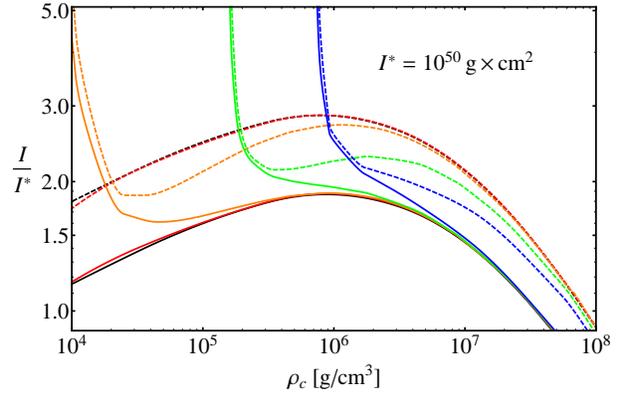} 
\caption{ Moment of inertia  versus central density. The legend is the same as in Fig.~\ref{fig:MR} (colour online).}\label{fig:Iphysrho} 
\end{figure}

\begin{figure}
\includegraphics[width=8.5cm]{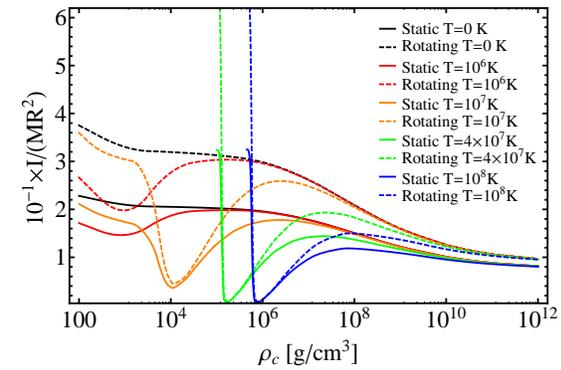} 
\caption{ Normalized moment of inertia  versus central density (colour online).}\label{fig:IMRrho} 
\end{figure}

Fig.~\ref{fig:Iphysrho}  shows the moment of inertia as a function of the central density for both static and rotating, cold and hot WDs.  In the static case the moment of inertia of  hot WDs is larger than for  cold ones, in the entire range of the central density. 
This was expected as hotter WDs with similar masses will be larger in size than colder ones. However, for rotating WDs the situation is slightly different as hotter (larger in size) WDs cannot rotate faster than colder (smaller in size) ones. This effect becomes more evident starting from the value of the central density $10^6$g/cm$^3$. Consequently, because of the rotation, the moment of inertia of hotter WDs will be smaller than that of colder ones.
For the normalized moment of inertia this effect is also valid in the static case as $MR^2$ goes up faster than the moment of inertia when temperature increases and the EoS becomes softer, for further details see Fig.\ref{fig:IMRrho}.

\begin{figure}
\includegraphics[width=8.5cm]{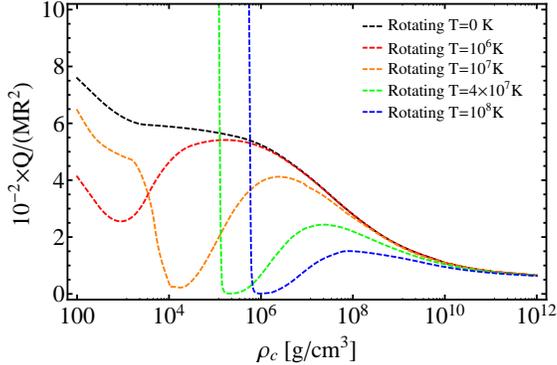} 
\caption{Normalized quadrupole moment  versus central density (colour online).}\label{fig:QMRrho} 
\end{figure}

The normalized quadrupole moment is shown as a function of the central density in Fig.~ \ref{fig:QMRrho}. The effect of the temperature is considerably small in the range of densities higher than $10^{10}$ g/cm$^3$. However, as the density diminishes the temperature plays a more important role, leading to a nonlinear behaviour of the analysed quantities. For values of the central density lower than $10^{10}$ g/cm$^3$, the quadrupole moment strongly depends on the temperature, but in general it increases in value for less massive stars. Within the approximate interval $\rho_c \in [10^4,10^6]$ g/cm$^3$ and for specific  values of the temperature, the quadrupole moment drastically decreases, indicating a trend towards spherical symmetry. 

\begin{figure}
\includegraphics[width=8.5cm]{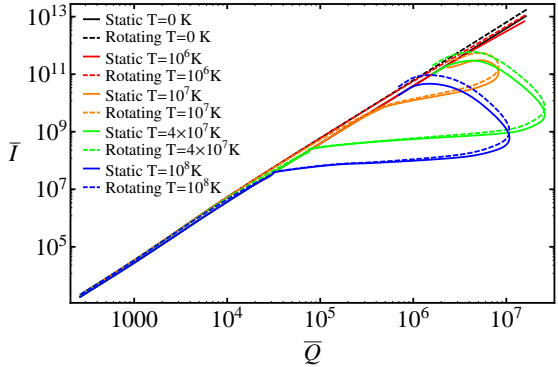} 
\caption{Dimensionless moment of inertia versus dimensionless quadrupole moment (colour online).}\label{fig:QbarIbar} 
\end{figure}

The above results show that temperature can play a very important role in the determination of the physical properties of WDs. Moreover,  
at first glance it seems that the moment of inertia and the quadrupole moment correlate. However, a deeper analysis shows a discrepancy. 
In Fig.\ref{fig:QbarIbar}, we plot the dimensionless moment of inertia $\bar{I}=(c^4I)/(G^2M^3)$ against the dimensionless quadrupole moment $\bar{Q}=(c^2Q)/(J^2/M)$, where $I$ is the physical moment of inertia, $Q$ is the physical mass quadrupole moment, $M$ is the static mass and $J$ is the angular momentum of the WD. For the 
degenerate case, $T=0$, we corroborate the $\bar I -\bar Q$ relation established previously by \citet{bosh2017}. As the temperature increases towards the range of realistic values the  $\bar I -\bar Q$ relation clearly breaks down. As the temperature increases, the break point 
moves towards the region of lower quadrupole moment. This proves that the  $\bar I -\bar Q$ is no longer valid in the case of hot WDs.   

\begin{figure}
\includegraphics[width=8.5cm]{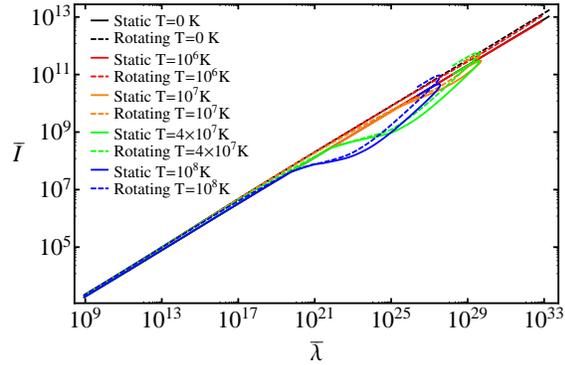} 
\caption{Dimensionless moment of inertia versus dimensionless tidal Love number (colour online).}\label{fig:IbarLbar} 
\end{figure}

\begin{figure}
\includegraphics[width=8.5cm]{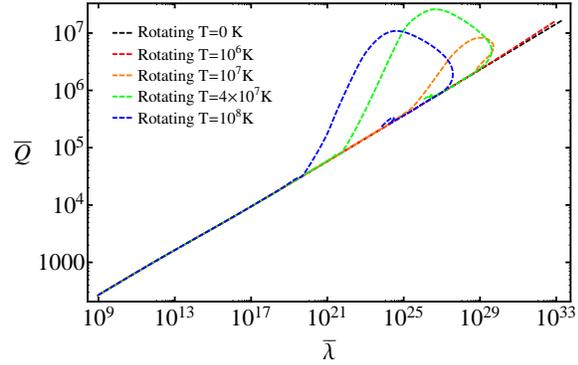} 
\caption{Dimensionless quadrupole moment versus dimensionless tidal Love number (colour online).}\label{fig:QbarLbar} 
\end{figure}

We investigate the $I$-Love-$Q$ relations in Figs.~\ref{fig:IbarLbar} and \ref{fig:QbarLbar}. As the temperature is taken into account, the non-validity of these relations becomes evident. We also see that as the temperature goes up, the breaking occurs at lower values of the 
dimensionless tidal Love number $\bar\lambda=(c^{10}\lambda)/(G^4M^5)$. 

Notice that in this approximation the moment of inertia can be expressed as the sum of a static
plus a rotational component, whereas the quadrupole moment has only a rotational component (for
the details of this decomposition, see \cite{bosh2016}). Therefore, although in Fig.~\ref{fig:QbarIbar} we use
$\bar Q$ as a parameter for the central density also in the static case, it does not mean that
there is a static quadrupole moment. Also for this reason, in Fig.~\ref{fig:QbarLbar} only the rotational
component of the quadrupole moment is plotted.
For the sake of clarity, we present in Table ~\ref{tab:t5} the numerical values for the $I$-Love-$Q$ relations in terms 
of the central density for zero temperature white dwarf stars.
Fig. ~\ref{fig:QbarIbar} and Fig.~\ref{fig:IbarLbar} illustrate the behavior of $\bar{I}$ and $\bar{I}+\Delta\bar{I}$ as  
functions of  $\bar{Q}$, where $\bar{Q}$ serves as a parameter for $\rho$, and $\bar{\lambda}$, respectively.
Finally, Fig.~\ref{fig:QbarLbar} represents $\bar{Q}$ as a function of  $\bar{\lambda}$.

\begin{table}

\centering
\caption{ $I$-Love-$Q$ relations for white dwarf stars with $T=0$ K. Here $\rho$ is the central density, 
$\bar{I}$ is the dimensionless moment of inertia for static configurations, 
$\bar{I}+\Delta\bar{I}$ is the dimensionless moment of inertia for rotating configurations,  $\bar{Q}$ is the dimensionless quadrupole moment for rotating configurations and $\bar{\lambda}$ is the dimensionless tidal Love number for static configurations.}
{\footnotesize
{\begin{tabular}{c c c c c c}

\hline
$\rho$ (g/cm$^3$) & $\bar{I}$ & $\bar{I}+\Delta\bar{I}$ & $\bar{Q}$& $\bar{\lambda}$ \\
\hline

 $10^2$      & 1.0$\times10^{13}$ & 1.7$\times10^{13}$ & 1.6$\times10^{7}$ & 1.7$\times10^{33}$ \\
 $10^3$      & 4.8$\times10^{11}$ & 7.6$\times10^{11}$ & 3.5$\times10^{6}$ & 8.2$\times10^{29}$ \\ 
 $10^4$      & 2.3$\times10^{10}$ & 3.5$\times10^{10}$ & 7.6$\times10^{5}$ & 4.0$\times10^{26}$ \\ 
 $10^5$      & 1.1$\times10^{9}$  & 1.7$\times10^{9}$  & 1.7$\times10^{5}$ & 2.2$\times10^{23}$ \\
 $10^6$      & 6.7$\times10^{7}$  & 1.0$\times10^{8}$  & 4.2$\times10^{4}$ & 1.9$\times10^{20}$ \\
 $10^7$      & 6.2$\times10^{6}$  & 9.1$\times10^{6}$  & 1.3$\times10^{4}$ & 4.9$\times10^{17}$ \\
 $10^8$      & 9.3$\times10^{5}$  & 1.3$\times10^{6}$  & 5.2$\times10^{3}$ & 4.5$\times10^{15}$ \\
 $10^9$      & 1.8$\times10^{5}$  & 2.4$\times10^{5}$  & 2.4$\times10^{3}$ & 8.2$\times10^{13}$ \\
 $10^{10}$   & 3.9$\times10^{4}$  & 4.9$\times10^{4}$  & 1.2$\times10^{3}$ & 1.8$\times10^{12}$ \\
 $10^{11}$   & 8.5$\times10^{3}$  & 1.0$\times10^{4}$  & 5.6$\times10^{2}$ & 4.0$\times10^{10}$ \\

\hline

\end{tabular}\label{tab:t5}}
 }
\end{table}

From the above results, we conclude that the $I$-Love-$Q$ relations proposed by \citet{yagi2013} for relativistic objects are not true for hot white dwarf stars, even in the framework of Newtonian gravity. The universality is thus lost for larger values of the moment of inertia, quadrupole moment and tidal Love number. In the region of smaller values of these parameters, which corresponds to the regime of larger densities when the degeneracy sets in, the behaviour is almost universal as it was shown by \citet{bosh2017}. 

The non-validity of the $I-$Love$-Q$ relations was also found in other studies. For instance, the breakdown of these relations was found by \citet{doneva2014} for rapidly rotating NSs and QSs, although in the slow-rotation approximation and at fixed rotational frequencies, one can still find roughly EoS-independent relations. Similar results have been obtained by \citet{pappas2014}. 
The $I$-$Q$ relations for arbitrarily fast rotating NSs were also considered by \citet{chak2014}, where it was found that the relations can be still universal among various EoSs for constant values of certain dimensionless parameters characterizing the magnitude of the rotation.
However, it was demonstrated by \citet{haskell2014} that the universality of the relations is lost in the presence of huge magnetic fields in NSs with rotation period larger than 10 seconds and magnetic fields larger than 10$^{12}$G. In addition, \citet{yagi2014} showed that the universality is also lost for non-compact objects when their opacity was varied.
Furthermore, the phase of the proto-NS life, including the effects of both rotation and finite temperatures, was studied by 
\citet{martinon2014}. It was shown that the $I$-Love-$Q$ relations are violated in the first second of life, but they are satisfied as soon as the entropy gradients smooth out. 
Recently, it was found that the $I$-$Q$ universality is broken when thermal effects become important, independently of the presence of entropy gradients \citep{marques2017}.

\section{Conclusions}

We numerically integrated the underlying differential equations in order to determine the structure of slowly and rigidly rotating classical WDs in hydrostatic equilibrium. In particular, using the Chandrasekhar EoS,  the relations for the  mass,  radius,  moment of inertia, and quadrupole moment were established  as  functions of the central density and temperature.
All these quantities play a crucial role in describing the equilibrium configurations of uniformly rotating main sequence stars as well as massive stars. In particular, we proved that the temperature affects the behavior of all the physical parameters, especially in the region of realistic temperature values. In addition, we calculated the tidal Love number and investigated the $I$-Love-$Q$ relations for rotating WDs. 
 
It turned out that the $I$-Love-$Q$ relations are not universal even within the same EoS when the finite temperature effects are taken into account. This is probably due to the fact that the EoS is not longer barotropic when the thermal effects are included, i.e., the pressure not only depends on the density, but also on the temperature. 
In a related work by \citet{lau2017}, it was found that the universality of the $I$-Love relation is broken when the elastic properties of crystalline quark matter are accounted for, i.e., the universality is observed only in perfect fluid compact objects (at zero temperature without magnetic fields). 

In view of their astrophysical relevance it would be interesting to investigate the validity of the  $I$-Love-$Q$ relations for WDs with different nuclear composition and magnetic field intensity \citep{malheiro2012, coelho2014, coelho2014b, lobato2016, alvear2017, alvear2017b, alvear2018}. This will be the issue of future studies.

\section*{Acknowledgement}
\medskip
\noindent
The authors thank anonymous referee for constructive suggestions and comments. This work was supported in part by UNAM-DGAPA-PAPIIT, Grant No. IN111617 and by Nazarbayev University Faculty Development Competitive Research Grants: Quantum gravity from outer space and the search for new extreme astrophysical phenomena, Grant No. 090118FD5348. K.B. thanks Jorge Rueda for fruitful discussions and acknowledges the ICRANet for hospitality, the MES of the RK for partial support, Program IRN: BR05236494.

\bibliographystyle{mn2e}

\end{document}